\documentclass[aps, prd, superscriptaddress, preprintnumbers, showpacs]{revtex4}

\usepackage{graphicx}
\usepackage{amsmath}
\usepackage{amssymb}
\usepackage{color}
\usepackage{bm}

\newcommand{\be}{\begin{equation}}
\newcommand{\ee}{\end{equation}}
\newcommand{\ba}{\begin{eqnarray}}
\newcommand{\ea}{\end{eqnarray}}

\newcommand{\tr}{\,\mbox{tr}}

\definecolor{purple}{rgb}{0.8,0,0.6}

\DeclareMathOperator{\arcsinh}{arcsinh}

\begin{document}

\title
{Gap generation in Weyl semimetals in a model with local four-fermion interaction}
\date{\today}


\author{P.O. Sukhachov}
\affiliation{Department of Physics, Taras Shevchenko National Kiev University, Kiev, 03680, Ukraine}



\setcounter{page}{1}%

\begin{abstract}
We study the gap generation in Weyl semimetals in a model with local four-fermion interaction. It is shown that there exists a critical value of
coupling constant separating the symmetric and broken symmetry phases, and the corresponding phase diagram is described. The gap
generation in a more general class of Weyl materials with small bare gap is studied, and the quasiparticle energy spectrum
is determined. It is found that, in this case the dynamically generated gap leads to the additional splitting of the quasiparticle energy bands.
\end{abstract}


\maketitle

\section{Introduction}
\label{1}

The discovery of new materials with unique quantum-mechanical properties is crucial for the progress in condensed matter physics.
Recently such new materials as topological insulators, Dirac semimetals, and Weyl semimetals attracted the attention of the condensed
matter community and moved at the forefront of theoretical and
experimental studies \cite{Zhang-Rev, Hasan-Rev, Ando-Rev}. Remarkable properties of these two-dimensional (2D) and 3D materials
are connected with the unusual properties of their low energy quasiparticle excitations, which are described by the Dirac or Weyl equation.
Since 3D massless Dirac fermions can be represented as two copies of Weyl fermions of opposite chirality, Weyl fermions can be considered as
the most elementary building blocks of these 3D materials. It is important to note that while two Weyl nodes
for every particle (except neutrinos which are perhaps only left-handed fermions) in the elementary particle physics are located at $\mathbf{k} = 0$ forming thus a Dirac
fermion, Weyl nodes in condensed matter physics are, in general, located at different points in the momentum space.

As is well known, graphene is a 2D Dirac semimetal. Consequently, 3D Dirac semimetals may be considered as 3D analogues of graphene.
The first historically known 3D Dirac material is bismuth \cite{Cohen, Wolff, Falkovskii, Edelman} whose electron states near
the L point in the Brillouin zone are described by 3D Dirac massive equation with sufficiently large Dirac mass. It is possible to decrease
Dirac mass by doping Bi with antimony. As the antimony concentration reaches $x\approx0.03$, alloy $\mathrm{Bi}_{1-x}\mathrm{Sb}_x$ transforms into a Dirac
semimetal with massless Dirac point, realizing thus a 3D analogue of graphene. Using the \emph{ab-initio} calculations and effective model
analysis, it was further theoretically suggested in Refs.~\cite{WangSun, WangWeng} that $\mathrm{Na}_3\mathrm{Bi}$, $\mathrm{K}_3\mathrm{Bi}$, $\mathrm{Rb}_3\mathrm{Bi}$, and
$\mathrm{Cd}_3 \mathrm{As}_2$ are 3D Dirac semimetals. By investigating the electronic structure with angle resolved photoemission spectroscopy, 3D Dirac fermions
were experimentally discovered in $\mathrm{Na}_3\mathrm{Bi}$ in Ref.\cite{Liu} and $\mathrm{Cd}_3 \mathrm{As}_2$ in Refs.~\cite{Neupane, Borisenko}. As to the Weyl semimetals,
the recent observation of negative magnetoresistivity in $\mathrm{Bi}_{0.97}\mathrm{Sb}_{0.03}$ provided an experimental evidence for the
existence of Weyl fermions Ref.~\cite{Kitaura}. To obtain a Weyl semimetal from a Dirac semimetal, one must break either
time reversal or inversion symmetry. This can be done, for example, by applying an external magnetic field. As a result, the 3D Dirac point splits
into two Weyl nodes of opposite chiralities. A good example of the dynamical transformation of a Dirac semimetal into a Weyl one is
given by the dynamical generation of the chiral shift parameter considered in Ref.~\cite{engineering}.

Since the Coulomb interaction is not screened in Weyl semimetals due to the vanishing of the density of states at the Fermi surface,
the electron-electron interactions in these materials are very important and may lead to the dynamical chiral symmetry breaking, which is
connected with the dynamical gap generation due to the pairing of electrons and holes with different chiralities. In this paper, we consider
the dynamical chiral symmetry breaking in Weyl semimetals in a model with local four-fermion interaction with regard for a small bare gap for
quasiparticles. The gap generation in Weyl semimetals in the absence of bare gap was previously studied in
Refs.~\cite{Wang, Wei, Yang}.

This paper is organized as follows. In Sec.~\ref{2} we introduce the model and set up the notations. The gap equation for case of zero bare gap is derived, solved and the dependence of the gap on the interaction strength and the momentum space separation between the
Weyl nodes is determined in Sec.~\ref{3}. The more general case of a nonzero bare gap is considered in Sec.~\ref{4}. Using the perturbation theory, we derived and solved gap equations. The energy spectrum was obtained and described. The results are summarized, and conclusions are given in Sec.~\ref{5}. For convenience, throughout this
paper, we set $\hbar=1$.

\vspace{5mm}

\section{Model}
\label{2}

We begin our study by considering the following low-energy Hamiltonian (see, Ref.~\cite{engineering}):
\begin{equation}
H^{\rm (W)}=H^{\rm (W)}_{0}+H_{\rm int},
\label{Hamiltonian-model-Weyl}
\end{equation}
where
\begin{eqnarray}
H^{\rm (W)}_{0}=-\int d^3r \psi^{\dagger} (\mathbf{r}) \left(\begin{array}{cc}
v_F\bm{\sigma}\cdot(i\bm{\nabla}+\mathbf{b}_0)  & \Delta_0\\
\Delta_0 & v_F\bm{\sigma}\cdot(-i\bm{\nabla}+\mathbf{b}_0)\\
\end{array}
\right)\psi(\mathbf{r})
\label{free-Hamiltonian}
\end{eqnarray}
is the Hamiltonian of the free theory and $\Delta_0$ is the bare gap parameter. This Hamiltonian describes two Weyl nodes
of opposite chiralities separated by the vector $2\mathbf{b}_0$ in the momentum space. The opposite chiralities of Weyl nodes are required by the
Nielsen--Ninomiya theorem \cite{NN}. Following Refs.~\cite{engineering,Gorbar:2009bm,Gorbar:2011ya}, we call $\mathbf{b}_0$ the bare chiral
shift parameter. Other notations: $v_F$ is the
Fermi velocity, and $\bm{\sigma}=(\sigma_x,\sigma_y,\sigma_z)$ are Pauli matrices associated with the band degrees of freedom
\cite{Burkov3,engineering}. In the general case, the interaction Hamiltonian $H_{\rm int}$
describes the Coulomb interaction, i.e.,
\begin{equation}
H_{\rm int} = \frac{1}{2}\int d^3rd^3r^{\prime}\,\psi^{\dagger}(\mathbf{r})\psi(\mathbf{r})U(\mathbf{r}-\mathbf{r}^{\prime})
\psi^{\dagger}(\mathbf{r}^{\prime})\psi(\mathbf{r}^{\prime}).
\label{int-Hamiltonian}
\end{equation}
In order to simplify our calculations, we will use a model with a contact four-fermion interaction
\begin{equation}
U(\mathbf{r}) = \frac{e^2}{\kappa |\mathbf{r}|} \rightarrow g\, \delta^3(\mathbf{r}),
\label{model-interaction}
\end{equation}
where $g$ is a dimensionful coupling constant. As we will see, this model interaction
should be sufficient for the general qualitative description of the gap generation in Weyl semimetals.
Before proceeding further with the analysis, it is convenient to introduce the four-dimensional Dirac matrices in the chiral representation:
\begin{eqnarray}
\gamma^0 = \left( \begin{array}{cc} 0 & -I\\ -I & 0 \end{array} \right),\,\,\,\,\,
\bm{\gamma} = \left( \begin{array}{cc} 0& \bm{\sigma} \\  - \bm{\sigma} & 0 \end{array} \right), \,\,\,\,\, \gamma^5 \equiv i\gamma^0\gamma^1\gamma^2\gamma^3 = \left( \begin{array}{cc} I & 0\\ 0 & -I \end{array} \right),
\label{Dirac-matrices}
\end{eqnarray}
where $I$ is the two-dimensional unit matrix. Using the Eq.~(\ref{model-interaction}) and Eq.~(\ref{Dirac-matrices}), we can rewrite the full Hamiltonian Eq.~(\ref{Hamiltonian-model-Weyl}) as follows:
\begin{eqnarray}
H^{\rm (W)}=\int d^3r \psi^{\dagger} (\mathbf{r})\left(-iv_F\gamma_0(\bm{\gamma}\cdot\bm{\nabla})+ v_F\gamma_0\gamma_5(\bm{\gamma}\cdot\mathbf{b}_0)+\gamma_0\Delta \right)
\psi(\mathbf{r}) +\frac{g}{2}\int d^3r\,\psi^{\dagger}(\mathbf{r})\psi(\mathbf{r})\psi^{\dagger}(\mathbf{r})\psi(\mathbf{r}).
\label{full-Hamiltonian-TI}
\end{eqnarray}
\vspace{5mm}

\section{Gap equation in Weyl semimetals without bare gap}
\label{3}

\subsection{Derivation of the gap equation}

In this section, we derive the gap equation in Weyl semimetals using the Cornwall--Jackiw--Tomboulis formalism \cite{CJT}.
The Cornwall--Jackiw--Tomboulis effective action in the first order of the perturbation theory takes the form:
\begin{eqnarray}
\Gamma(G) = -i\,\mathrm{Tr}[\mathrm{Ln}G^{-1}+S^{-1}G -1]+\frac{g}{2}\int d^4r \left(\mathrm{tr}\big[G(r,r)G(r,r)\big] -\mathrm{tr}\big[G(r,r)\big]\mathrm{tr}\big[G(r,r)\big] \right),
\label{CJT-LR-1}
\end{eqnarray}
where $G$ is the full fermion propagator, and $S$ is the free fermion propagator. The trace and the logarithm in the first term on the
right-hand side of the above equation are taken in the functional sense. The Schwinger-Dyson equation for the fermion propagator
determines extrema of the Cornwall--Jackiw--Tombolulis effective action and is given by
\begin{equation}
G^{-1}(r, r') = S^{-1}(r, r')+ig\delta^{(4)}(r-r') (G - \tr[G]),
\label{CJT-LR-2}
\end{equation}
where the trace is taken over spinor indices. The inverse free fermion propagator is given by
\begin{eqnarray}
iS^{-1}(r,r^\prime) = \left(i\partial_t +iv_F\gamma_0(\bm{\gamma}\cdot\bm{\nabla})- v_F\gamma_0\gamma_5(\bm{\gamma}\cdot\mathbf{b}_0)\right)\delta^{4}(r-r^{\prime}),
\label{sinverse}
\end{eqnarray}
and an ansatz for the inverse full fermion propagator is given by
\begin{eqnarray}
iG^{-1}(r, r^{\prime}) = \left(i\partial_t +iv_F\gamma_0(\bm{\gamma}\cdot\bm{\nabla})-
v_F\gamma_0\gamma_5(\bm{\gamma}\cdot\mathbf{b}) -\gamma_0\Delta e^{-2i(\mathbf{b}^{\prime}\cdot\mathbf{r})\gamma_5} \right)\delta^{4}(r-
r^{\prime}),
\label{green-function-full}
\end{eqnarray}
where $\mathbf{b}$ is a renormalized chiral shift and $\Delta e^{-2i(\mathbf{b}^{\prime}\cdot\mathbf{r})\gamma_5}$ is the general form
of the gap term, which can be understood as the chiral charge density wave order parameter. This form of chiral condensation, where
fermions (electrons) and antifermions (holes) are paired in a state with total momentum $2\mathbf{b}^{\prime}$, is reminiscent of the
Larkin--Ovchinnikov--Fulde--Ferrell (LOFF) \cite{LO,FF} state of pairing between electrons with nonzero total momentum in the theory of
superconductivity. Obviously, this phase can be eliminated by the chiral transformation
\begin{equation}
iG^{-1}(r,r^\prime)= e^{i(\mathbf{b}^{\prime}\cdot\mathbf{r})\gamma_5}i\bar{G}^{-1}(r,r^\prime)
e^{-i(\mathbf{b}^{\prime}\cdot\mathbf{r}^{\prime})\gamma_5},
\label{chiral-transformation}
\end{equation}
where
\begin{eqnarray}
i\bar{G}^{-1}(r, r^{\prime}) = \left(i\partial_t +iv_F\gamma_0(\bm{\gamma}\cdot\bm{\nabla})-
v_F\gamma_0\gamma_5(\bm{\gamma}\cdot\bar{\mathbf{b}})- \gamma_0\Delta \right)\delta^{4}(r-
r^{\prime})
\label{ginverse-separated}
\end{eqnarray}
is the inverse fermion propagator with conventional Dirac mass without chiral phase and $\bar{\mathbf{b}}=\mathbf{b}-\mathbf{b}^{\prime}$. In the momentum space, Eq.~(\ref{ginverse-separated}) takes the following form
\begin{eqnarray}
i\bar{G}^{-1}(\omega, \mathbf{k})= \left(
\begin{array}{cc} \omega-v_F\bm{\sigma}\cdot(\mathbf{k}-\bar{\mathbf{b}})  & \Delta\\
\Delta & \omega+v_F\bm{\sigma}\cdot(\mathbf{k}+\bar{\mathbf{b}}) \end{array}
\right).
\label{green-function-full-momentum}
\end{eqnarray}
Multiplying the Schwinger--Dyson equation (\ref{CJT-LR-2}) by $e^{-i(\mathbf{b}^{\prime}\cdot\mathbf{r})\gamma_5}$ from the left and
$e^{i(\mathbf{b}^{\prime}\cdot\mathbf{r}^{\prime})\gamma_5}$ from the right, we obtain the following equation:
\begin{equation}
i\bar{G}^{-1}(\omega, \mathbf{k}) = i\bar{S}^{-1}(\omega, \mathbf{k})- g(\bar{G} - \tr[\bar{G}]),
\label{CJT-LR-2-momentum}
\end{equation}
where $\bar{S}^{-1}(\omega, \mathbf{k})$ coincides with the inverse free propagator with $\mathbf{b}_0$ replaced by relative chiral shift
$\bar{\mathbf{b}}_0=\mathbf{b}_0-\mathbf{b}^{\prime}$.
Multiplying Eq.~(\ref{CJT-LR-2-momentum}) by $\gamma^{0}\gamma^{5}\bm{\gamma}$ and taking trace, we obtain the following equation for the chiral
shift parameter:
\begin{equation}
\bar{\mathbf{b}}=\bar{\mathbf{b}}_0+\frac{g}{4v_F} \tr{\left[\gamma^{0}\gamma^{5}\bm{\gamma}\bar{G}\right]}.
\label{momentum-shift}
\end{equation}
Further, multiplying Eq.~(\ref{CJT-LR-2-momentum}) by $\gamma_0$ and taking trace, we find the gap equation
\begin{equation}
\Delta=\frac{g}{4} \tr\left[\gamma^0G\right].
\label{mass-gap}
\end{equation}
Inverting Eq.~(\ref{green-function-full-momentum}), we obtain the full fermion propagator
\begin{eqnarray}
\hspace{-4mm}&&i\bar{G}(\omega, \mathbf{k})N = \omega K_{0} +v_F\left(K_0\mathbf{k}-2v_F^2(\mathbf{k}\cdot\bar{\mathbf{b}})\bar{\mathbf{b}} \right) \gamma^0\bm{\gamma}+ v_F\left[2v_F^2(\mathbf{k}\cdot\bar{\mathbf{b}})\mathbf{k}-(K_0-2\Delta^2)\bar{\mathbf{b}} \right]
\gamma^5\gamma^0\bm{\gamma}+ \nonumber\\
\hspace{-4mm}&&+2v_F^2(\mathbf{k}\cdot\bar{\mathbf{b}})\omega\gamma^{5}+\Delta\left(K_0-2v_F^2\bar{\mathbf{b}}^2\right) \gamma^{0}+ 2iv_F^2\Delta([\bar{\mathbf{b}}\times\mathbf{k}]\cdot\bm{\gamma})-2\omega\Delta v_F \gamma^{5}
(\bm{\gamma}\cdot\bar{\mathbf{b}}),
\label{green-function-LR}
\end{eqnarray}
where $K_{0}=v_F^2\left(\mathbf{k}^2+\bar{\mathbf{b}}^2\right)+\Delta^2-\omega^2$ and $N=K_0^2-4v_F^2\left(\Delta^2\bar{\mathbf{b}}^2+v_F^2(\mathbf{k}\cdot\bar{\mathbf{b}})^2\right)$. We can integrate over the frequency on the right-hand
side of Eqs.~(\ref{momentum-shift}) and (\ref{mass-gap}). These integrals have a similar structure and can be easily calculated
\begin{eqnarray}
 \int d\omega \frac{A_1+A_2\omega^2}{(\omega^2+W_1)(\omega^2+W_2)}= \frac{\pi}{\sqrt{W_1}+\sqrt{W_2}}\left(A_2+\frac{A_1}{\sqrt{W_1W_2}}\right),
\label{gap-integrals-01}
\end{eqnarray}
where $W_{1, 2}=\left(\Delta^2+v_F^2(\mathbf{k}^2+\bar{\mathbf{b}}^2)\right)\mp \sqrt{K_0^2-N}$.
Thus, we obtain the following
system of equations:
\begin{eqnarray}
1=g\int\frac{d^3\mathbf{k}}{(2\pi)^4}\frac{\pi }{\sqrt{W_1}+\sqrt{W_2}} \left(1+\frac{v_F^2(\mathbf{k}^2-\bar{\mathbf{b}}^2)+\Delta^2}{\sqrt{W_1W_2}}\right),
\label{mass-gap-02}
\end{eqnarray}
\begin{eqnarray}
\bar{\mathbf{b}}=\bar{\mathbf{b}}_{0}-g\int\frac{d^3\mathbf{k}}{(2\pi)^4}\frac{\pi }{\sqrt{W_1}+\sqrt{
W_2}} \left(\bar{\mathbf{b}}+\frac{\bar{\mathbf{b}}\left(v_F^2(\bar{\mathbf{b}}^2+\mathbf{k}^2)-\Delta^2\right)-2\mathbf{k}v_F^2(\bar{\mathbf{b}}\cdot\mathbf{k})}
{\sqrt{W_1W_2}}\right).
\label{momentum-shift-02}
\end{eqnarray}

\subsection{Solution with chiral phase}

In this case, fermions and antifermions are paired with non-zero total momentum. One can easily prove that $\bar{\mathbf{b}}=0$ if we choose $\mathbf{b}^{\prime}=\mathbf{b}_{0}$, which leads to $\bar{\mathbf{b}}_{0}=0$. Then Eq.~(\ref{mass-gap-02}) equals
\begin{eqnarray}
\hspace{-4mm}&&1=g\int\frac{d^3\mathbf{k}}{(2\pi)^4}\frac{\pi }{\sqrt{\Delta^2+v_F^2k^2}}= 4g\pi^2\int_0^{\Lambda}\frac{k^2dk}{(2\pi)^4}\frac{1}
{\sqrt{\Delta^2+v_F^2k^2}}= \nonumber\\
\hspace{-4mm}&&=\frac{g\Lambda^2}{8v_F\pi^2}\left(\sqrt{\left(\frac{\Delta}{v_F\Lambda}\right)^2+1}-\left(\frac{\Delta}{v_F\Lambda}\right)^2\arcsinh{\left(\frac{v_F\Lambda}{\Delta}\right)}\right),
\label{mass-gap-b0-01}
\end{eqnarray}
where $\Lambda=\frac{\pi}{a}$ is a momentum cutoff, and $a$ is the lattice spacing. Assuming that
 $\frac{\Delta}{v_F\Lambda}\ll1$, Eq.~(\ref{mass-gap-b0-01}) simplifies to the following one:
\begin{eqnarray}
\hspace{-4mm}&&\frac{1}{g}-\frac{1}{g_{cr}} \approx \frac{\Delta^2}{16v_F^3\pi^2}\left( 1+2\ln\left(\frac{\Delta}{2v_F\Lambda}\right) \right), \nonumber\\
\hspace{-4mm}&&g_{cr}=\frac{8\pi^2v_F}{\Lambda^2},
\label{mass-gap-b0-02}
\end{eqnarray}
where $g_{cr}$ is the critical value of coupling constant.

\begin{figure}[h]
\begin{center}
\includegraphics[width=0.6\textwidth]{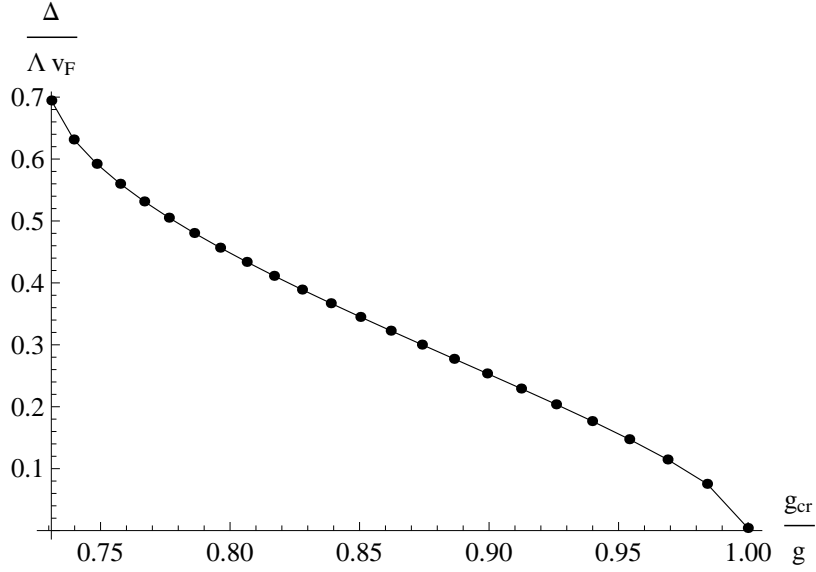}
\caption{Gap as a function of $g_{cr}/g$.}
\label{fig:b0}
\end{center}
\end{figure}

One can see from Eq.~(\ref{mass-gap-b0-02}) and Fig.~\ref{fig:b0} that the coupling constant $g$ must exceed a critical value $g_{cr}$
in order to produce non-trivial gap $\Delta$. Of course, there is also the trivial solution $\Delta=0$. To determine the solution with the
lowest energy, we will calculate the value of the Cornwall--Jackiw--Tomboulis effective action at its extrema, which gives the
energy of the system. After some calculations (for more details see Appendix~\ref{A3}), we find the energy density of the system
\begin{eqnarray}
{\cal E} =-\frac{\Lambda^4 v_F}{8\pi^2}\left[\sqrt{\left(\frac{\Delta}{v_F\Lambda}\right)^2+1}\left(2-\left(\frac{\Delta}{v_F\Lambda}\right)^2\right)+\left(\frac{\Delta}{v_F\Lambda}\right)^4\arcsinh{\left(\frac{v_F\Lambda}{\Delta}\right)} \right].
\label{effective-potential-phase-02}
\end{eqnarray}
Since ${\cal E}(\Delta\neq0)-{\cal E}(\Delta=0) < 0$ for $\frac{\Delta}{v_F\Lambda}<1$, a non-trivial solution is always more favorable as soon as it exists.
Our results coincide with those obtained in Refs.~\cite{Wang, Wei, Yang}.

\subsection{Phase diagram}

\begin{figure}[h]
\begin{center}
\includegraphics[width=0.5\textwidth]{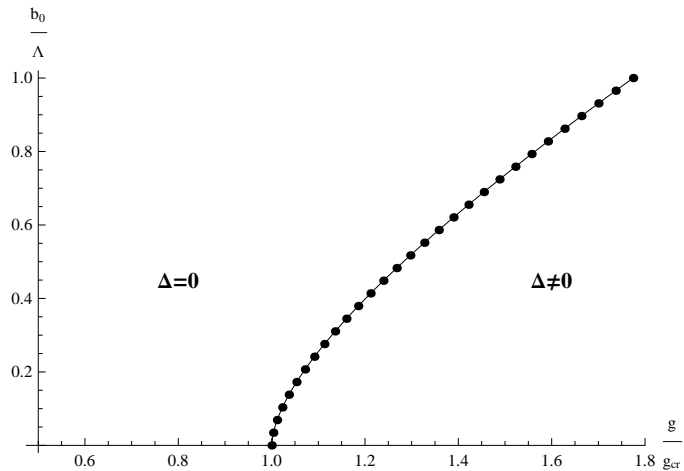}
\caption{Domain of existence of the Dirac phase.}
\label{fig:Phase-diagram-Dirac}
\end{center}
\end{figure}

\begin{figure}[h!]
\begin{center}
\includegraphics[width=0.5\textwidth]{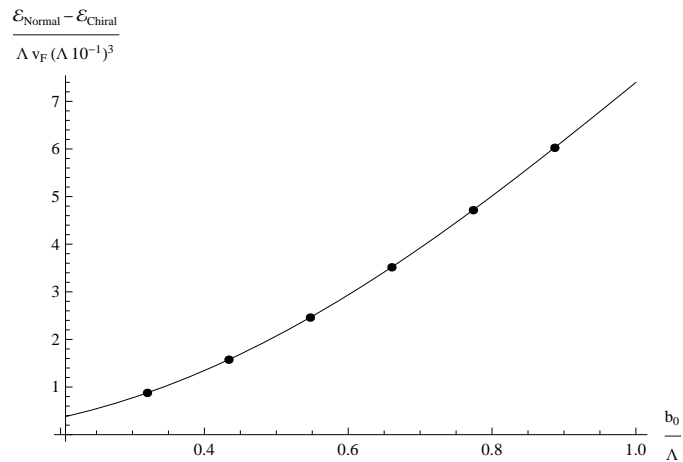}
\caption{Difference of energy densities ${\cal E}_{Normal}-{\cal E}_{Chiral}$ as a function of $b_0$.}
\label{fig:Energy-difference-Normal}
\end{center}
\end{figure}

\begin{figure}[h!]
\begin{center}
\includegraphics[width=0.5\textwidth]{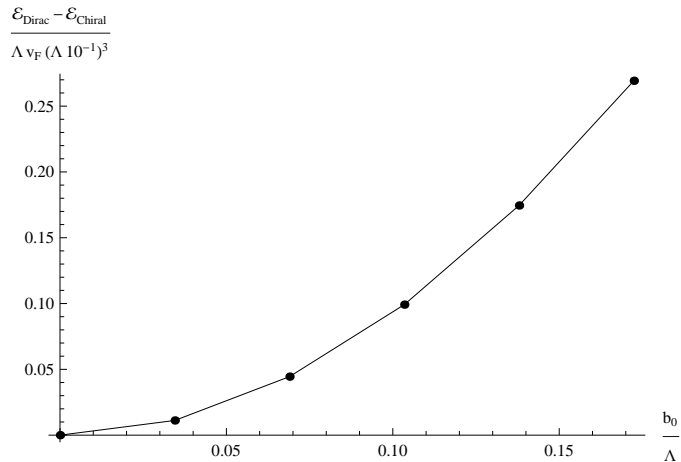}
\caption{Difference of energy densities  ${\cal E}_{Dirac}-{\cal E}_{Chiral}$ as a function of $b_0$.}
\label{fig:Energy-difference-Dirac}
\end{center}
\end{figure}

In this subsection we compare three different phases that can exist in the system. The solution with chiral phase was studied in the previous subsection. In the case of Dirac phase, fermions and antifermions are paired with zero total momentum that means $\mathbf{b}^{\prime}=0$. Thus, we have usual Dirac mass term $\gamma_0\Delta$ in Eq.~(\ref{green-function-full}), and there is no need in the chiral transformation Eq.~(\ref{chiral-transformation}). Moreover, there is the normal phase, where $\Delta=0$. For both normal and Dirac phases, Eqs.~(\ref{mass-gap-02}) and (\ref{momentum-shift-02}) retain their form, but with replacement $\bar{\mathbf{b}}\rightarrow\mathbf{b}$. Without any loss of generality, we can assume that $\mathbf{b}_0$ and $\mathbf{b}$ point in the $+z$ direction. Eqs.~(\ref{mass-gap-02}) and (\ref{momentum-shift-02}) were solved numerically by using \emph{Mathematica} and the iteration procedure with the following values of constants:
$v_F=3.5\times 10^5\, m/s$, $\Lambda=\frac{\pi}{a}=2.65\times10^9 \,m^{-1}$ (according to Ref.~\cite{Malik}, for
$\mathrm{Bi}_{0.88}\mathrm{Sb}_{0.12}$, $a=1.18 \,nm$). The domain of existence of the Dirac phase is plotted in Fig.~\ref{fig:Phase-diagram-Dirac}, where the Dirac phase exist to the right from the critical line separating the symmetric normal phase and the Dirac phase with broken symmetry.
To obtain the full phase diagram, it is important to compare the energy density of the chiral (or LOFF-like) phase ${\cal E} (b^{\prime}\neq0, \Delta\neq0)$, with the energy density of the normal ${\cal E} (\Delta=0)$ and Dirac ${\cal E} (b^{\prime}=0, \Delta\neq0)$ phases, using the expression for energy density given by Eq.~(\ref{effective-potential-b-02}) in Appendix~\ref{A3}. Further, in this subsection we will use $g=1.05\times g_{cr}$. Using Eqs.~(\ref{effective-potential-b-02}) and (\ref{green-function-LR}), we calculate the energy densities of these phases as functions of $b_0$. The difference of energy densities of the normal and chiral phases and the difference of energy densities of the Dirac and chiral phases are plotted in Figs.~\ref{fig:Energy-difference-Normal} and \ref{fig:Energy-difference-Dirac}, respectively.

\begin{figure}[h]
\begin{center}
\includegraphics[width=0.5\textwidth]{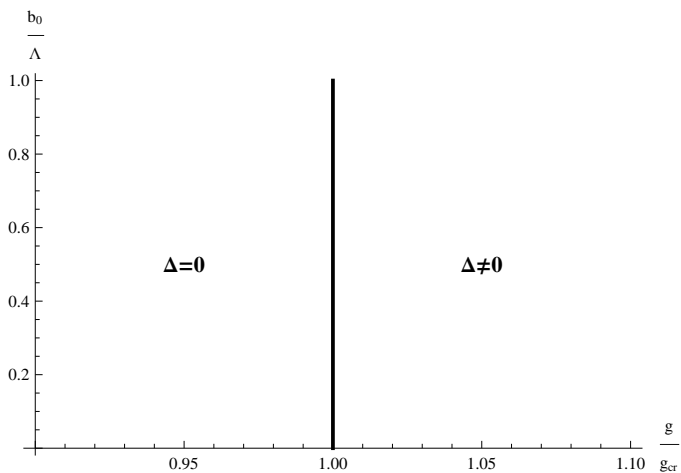}
\caption{Phase diagram of the system.}
\label{fig:Phase-diagram-LOFF}
\end{center}
\end{figure}

We found that the phase diagram of the system is simple. For $g>g_{cr}$, the chiral phase has lower energy compared to that of the normal and Dirac phases. The phase diagram of the system is plotted in Fig.~\ref{fig:Phase-diagram-LOFF}.

\section{Gap equation in Weyl semimetals with bare gap $\Delta_0$}
\label{4}

\subsection{Derivation of the gap equation}

Let us consider a more general case of Weyl semimetal-like materials with bare gap $\Delta_0$. This case is of interes from the theoretical as
well as experimental viewpoints. For example, quasiparticle excitations in $\mathrm{Bi}_{1-x}\mathrm{Sb}_x$ are described by massless Dirac fermions only at one
point $x=0.03$, otherwise, a non-zero mass for Dirac quasiparticles is present. Using Eq.~(\ref{full-Hamiltonian-TI}), it is easy to
obtain the inverse free propagator
\begin{eqnarray}
iS^{-1}(r, r^{\prime}) = \left(i\partial_t +iv_F\gamma_0(\bm{\gamma}\cdot\bm{\nabla})- v_F\gamma_0\gamma_5(\bm{\gamma}\cdot\mathbf{b}_0)-\gamma_0\Delta_0\right)\delta^{4}(r- r^{\prime}).
\end{eqnarray}
As to the inverse full fermion propagator, the generalization of ansatz (\ref{green-function-full-momentum}) to the case under
consideration is given by
\begin{eqnarray}
iG^{-1}_1(r, r^{\prime}) = \left(i\partial_t +iv_F\gamma_0(\bm{\gamma}\cdot\bm{\nabla})-
v_F\gamma_0\gamma_5(\bm{\gamma}\cdot\mathbf{b})- \gamma_0(\Delta_1+\Delta e^{-2i(\mathbf{b}^{\prime}\cdot\mathbf{r})\gamma_5})\right)\delta^{4}
(r- r^{\prime}).
\label{green-function-inverse-massive}
\end{eqnarray}
To proceed further with the Schwinger-Dyson equation (\ref{CJT-LR-2}), we must calculate the fermion propagator. However, due to
$\Delta_0$ and $\Delta_1$ terms in the inverse full fermion propagator (\ref{green-function-inverse-massive}), the chiral phase factor
$e^{-2i(\mathbf{b}^{\prime}\cdot\mathbf{r})\gamma_5}$ cannot be removed by the chiral transformation (\ref{chiral-transformation}).
Therefore, we cannot proceed as straightforwardly as in the Sec.~\ref{3}. Since we assume that $\Delta_0$ is small, and $\Delta_1$ is
proportional to $\Delta_0$, we use the perturbation theory in $\Delta_1$. We have
\begin{equation}
G_1(r, r^{\prime}) = G(r, r^{\prime})+\Delta_1F(r, r^{\prime}),
\label{green-function-massive}
\end{equation}
where $G(r, r^{\prime})$ is the fermion propagator, which corresponds to the case $\Delta_1=0$.
To find $F(r, r^{\prime})$, we use the equation
\begin{equation}
\int d^4r^{\prime} G^{-1}_1(r, r^{\prime})G_1(r^{\prime}, r^{\prime \prime}) = \delta^{4}(r - r^{\prime \prime}).
\label{orthogonality}
\end{equation}
In the first order in $\Delta_1$, we find
\begin{equation}
\gamma_0iG(r, r^{\prime \prime})+\int d^4r^{\prime} G^{-1}(r, r^{\prime})F(r^{\prime}, r^{\prime \prime})=0.
\label{orthogonality-01}
\end{equation}
It is convenient to factor out the chiral phase in the inverse full and full fermion propagators
\begin{eqnarray}
\hspace{-4mm}&&G^{-1}(r, r^{\prime}) = e^{x}\bar{G}^{-1}(r, r^{\prime})e^{-x^{\prime}}, \nonumber\\
\hspace{-4mm}&&G(r, r^{\prime}) = e^{x}\bar{G}(r,
r^{\prime})e^{-x^{\prime}},
\label{G-phases}
\end{eqnarray}
where $\bar{G}(r, r^{\prime})$ is the fermion propagator without the chiral phase, and
$x\equiv i(\mathbf{b}^{\prime}\cdot\mathbf{r})\gamma_5$,
$x^{\prime}\equiv i(\mathbf{b}^{\prime}\cdot\mathbf{r}^{\prime})\gamma_5$.
Multiplying  Eq.~(\ref{orthogonality-01}) by $-ie^{x^{\prime \prime \prime}}\bar{G}(r^{\prime \prime \prime}, r)e^{-x}$ and
integrating over $r$, we obtain
\begin{eqnarray}
\int d^4r e^{x^{\prime \prime \prime}}\bar{G}(r^{\prime \prime \prime}, r)e^{-x} \gamma_0 e^{x}\bar{G}(r, r^{\prime
\prime})e^{-x^{\prime \prime}}= iF(r^{\prime \prime \prime}, r^{\prime \prime}).
\label{G-new-02}
\end{eqnarray}
The Schwinger-Dyson equation~(\ref{CJT-LR-2}) takes the following form:
\begin{eqnarray}
\hspace{-4mm}&&i\bar{G}^{-1}(r, r^{\prime}) -\gamma_0\Delta_1\delta^{(4)}(r - r^{\prime})e^{2x^{\prime}}= \nonumber\\
\hspace{-4mm}&&= e^{-x}iS^{-1}(r, r^{\prime})e^{x^{\prime}}
- g\delta^{(4)}(r-r^{\prime}) \left(\bar{G}(r, r^{\prime})+ \Delta_1e^{-x}F(r, r^{\prime})e^{x^{\prime}} - e^{-x}\tr[G_1]e^{x^{\prime}}\right).
\label{CJT-New-01}
\end{eqnarray}
Multiplying Eq.~(\ref{CJT-New-01}) by $\gamma_0$ and taking trace, we have
\begin{eqnarray}
4\Delta_1\cos{2(\mathbf{b}^{\prime}\cdot\mathbf{r})}+4\Delta = 4\Delta_0\cos{2(\mathbf{b}^{\prime}\cdot\mathbf{r})}+ g\tr{[\bar{G}_{0}
(r, r)\gamma_0]}+g\Delta_1\tr{[e^{-x}F(r, r)e^{x}\gamma_0]}.
\label{Gap-new}
\end{eqnarray}
The equation for $\Delta$ is the same as in Sec.~\ref{3} and can be easily written in the explicit form. Now, we can proceed with the
$F$ term:
\begin{eqnarray}
\tr{[e^{-x}iF(r, r)e^{x}\gamma_0]} =  \tr{\big[ \int d^4r^{\prime} \bar{G}(r, r^{\prime})e^{-2x^{\prime}}
\gamma_0 \bar{G}(r^{\prime}, r) \gamma_0\big]}.
\label{Gap-part-New}
\end{eqnarray}
Expressing $\bar{G}(r, r^{\prime})$ through its Fourier transform in the momentum space and integrating over $r$, we find
\begin{eqnarray}
\hspace{-4mm}&&\tr{[e^{-x}iF(r, r)e^{x}\gamma_0]} = \tr\big[ \int\frac{\,d\omega_1 d^3\mathbf{k}_1}{(2\pi)^4}
\frac{\,d\omega_2 d^3\mathbf{k}_2}{(2\pi)^4} \bar{G}(\omega_1, \mathbf{k}_1) (2\pi)^4\delta(\omega_1-\omega_2) \delta(\mathbf{k}_1-\mathbf{k}_2-2\mathbf{b}^{\prime}\gamma_5) \times \nonumber\\
\hspace{-4mm}&&\times e^{it^{\prime}(\omega_1-\omega_2)-i\mathbf{r}(\mathbf{k}_2-\mathbf{k}_1)} \gamma_0 \bar{G}(\omega_2,
\mathbf{k}_2) \gamma_0\big],
\label{Gap-part-New-02}
\end{eqnarray}
where $\delta(\mathbf{k}_1-\mathbf{k}_2+2\mathbf{b}^{\prime}\gamma_5)$ is a matrix $4\times4$ which can be written as:
\begin{eqnarray}
\delta(\mathbf{k}_1-\mathbf{k}_2-2\mathbf{b}^{\prime}\gamma_5) = \frac{1+\gamma_5}{2}\delta(\mathbf{k}_1-\mathbf{k}_2-2\mathbf{b}^{\prime}) + \frac{1-\gamma_5}{2}\delta(\mathbf{k}_1-\mathbf{k}_2+2\mathbf{b}^{\prime}).
\label{delta-matrix}
\end{eqnarray}
Integrating over $\omega_1$ and $\mathbf{k}_1$, Eq.~(\ref{Gap-part-New-02}) can be rewritten as follows:
\begin{eqnarray}
\tr{[e^{-x}iF(r, r)e^{x}\gamma_0]} = \tr\big[ \int\frac{\,d\omega_2 d^3\mathbf{k}_2}{(2\pi)^4} \left( \bar{G}(\omega_2, \mathbf{k}_2+2\mathbf{b}^{\prime}) P_{+} + \bar{G}(\omega_2,\mathbf{k}_2-2\mathbf{b}^{\prime})P_{-} \right) e^{2x} \bar{G}(\omega_2, -\mathbf{k}_2)\big],
\label{Gap-part-New-04}
\end{eqnarray}
where $P_{\pm}=\frac{1\pm\gamma_5}{2}$. Eq.~(\ref{Gap-part-New-04}) gives contributions only with the $\cos{2(\mathbf{b}^{\prime}\cdot\mathbf{r})}$ and $\sin{2(\mathbf{b}^{\prime}\cdot\mathbf{r})}$ terms. The sine term is approximately by $8$ orders smaller then the leading cosine term and will be neglected. This term is related to the approximations that were used in the derivation of the gap equation. Thus, the gap equation
(\ref{Gap-new}) is equivalent to
the following system of equations:
\begin{eqnarray}
\hspace{-4mm}&&\Delta_1 = \Delta_0+\frac{g}{4} \Delta_1 \tr{[e^{-x}F(r, r)e^{x}\gamma_0]}, \nonumber\\
\hspace{-4mm}&&\Delta =  \frac{ig}{4} \tr{[-i\bar{G}(r, r)\gamma_0]}.
\label{Gap-new-system}
\end{eqnarray}
To obtain the equation for the chiral shift parameter we multiply Eq.~(\ref{CJT-New-01}) by $\gamma^{0}\gamma^{5}\bm{\gamma}$ and take trace
\begin{eqnarray}
\bar{\mathbf{b}}=\bar{\mathbf{b}}_{0} + \frac{g}{4}\tr{[\gamma^{0}\gamma^{5}\bm{\gamma}\bar{G}_{0}(r, r)]} +
\frac{g\Delta_1}{4}\tr{[\gamma^{0}\gamma^{5}\bm{\gamma}e^{-x}F(r, r)e^{x}]}.
\label{Chiral-shift-01}
\end{eqnarray}
Further,
\begin{eqnarray}
\tr{[\gamma^{0}\gamma^{5}\bm{\gamma}e^{-x^{\prime}}iF(r^{\prime}, r^{\prime})e^{x^{\prime}}]} =  \tr\big[\gamma^{5}\bm{\gamma}
\int\frac{\,d\omega_2 d^3\mathbf{k}_2}{(2\pi)^4} \left( \bar{G}(\omega_2, \mathbf{k}_2+2\mathbf{b}^{\prime}) P_{+} + \bar{G}(\omega_2,
\mathbf{k}_2-2\mathbf{b}^{\prime})P_{-} \right) e^{2x^{\prime}}\bar{G}(\omega_2, -\mathbf{k}_2) \big].
\label{Chiral-shift-New}
\end{eqnarray}
This term also can generate only the sine and cosine terms.  Therefore, in order to be consistent with the initial ansatz for the fermion propagator
Eq.~(\ref{green-function-massive}), we should neglect them in the equation for the chiral shift parameter. So, we have
\begin{equation}
\bar{\mathbf{b}}=\bar{\mathbf{b}}_{0} + \frac{g}{4}\tr{[\gamma^{0}\gamma^{5}\bm{\gamma}\bar{G}(r, r)]},
\label{Chiral-shift}
\end{equation}
which is equivalent to Eq.~(\ref{momentum-shift}). Further, using Eq.~(\ref{green-function-LR}) and performing Wick rotation, we can obtain the following equation for $\Delta_1$:
\begin{eqnarray}
\hspace{-4mm}&&\Delta_1 = \Delta_0+\int \frac{d\omega_{E} d^3\mathbf{k}}{(2 \pi)^4} g \Delta_1 \left( \frac{(v_F^2k^2 + \omega_{E}^2)^2-\Delta^4}{K_{1}(\omega_{E}, \mathbf{k}) K_{1}(\omega_{E}, \mathbf{k}-2\mathbf{b}^{\prime}) K_{1}(\omega_{E},
\mathbf{k}+2\mathbf{b}^{\prime})} - \right.\nonumber\\
\hspace{-4mm}&&\left.-\frac{4b^{\prime 2}_{z}v_F^2\left(
k^2v_F^2\cos{(2\theta)}-W^2+\Delta^2\right)}{K_{1}(\omega_{E}, \mathbf{k}) K_{1}(\omega_{E}, \mathbf{k}-2\mathbf{b}^{\prime}) K_{1}(\omega_{E},
\mathbf{k}+2\mathbf{b}^{\prime})}\right).
\label{system-new}
\end{eqnarray}
where $K_{1}(\omega_{E}, \mathbf{k})=v_F^2\mathbf{k}^2+\Delta^2+\omega_{E}^2$.
It is worth mentioning that equations for $\Delta$ and $\mathbf{b}$ given by Eqs.(\ref{Gap-new-system}) and (\ref{Chiral-shift})
coincide with Eqs.~(\ref{mass-gap-02}) and (\ref{momentum-shift-02}) in Sec.~\ref{3}.

\subsection{Solutions}

\begin{figure}[h]
\begin{center}
\includegraphics[width=0.62\textwidth]{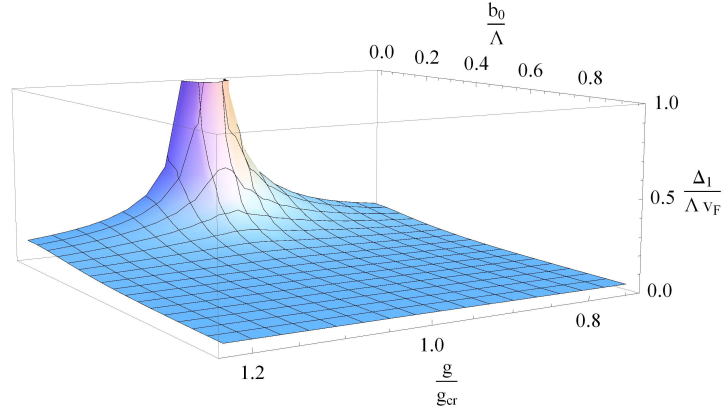}
\caption{Dependence of $\Delta_1$ on the coupling constant and the chiral shift parameter.}
\label{fig:Delta_Surface}
\end{center}
\end{figure}

Eq.~(\ref{system-new}) is solved numerically in the case of $\mathbf{b}^{\prime}=\mathbf{b}_{0}$ and $\mathbf{b}_{0}=\{0, 0, b_0\}$, by using \emph{Mathematica}. Further, we use the following values of constants:
$v_F=3.5\times 10^5\, m/s$, $\Lambda=\frac{\pi}{a}=2.65\times10^9 \,m^{-1}$ (according to Ref.~\cite{Malik}, for
$\mathrm{Bi}_{0.88}\mathrm{Sb}_{0.12}$, $a=1.18 \,nm$), and $\Delta_0 = 0.021\, eV$ (according to
Ref.~\cite{Malik}).
Numerical solutions of Eq.~(\ref{system-new}) are plotted in Fig.~\ref{fig:Delta_Surface}.

\subsection{Quasiparticle energy spectrum}

\begin{figure*}
\begin{minipage}[h!]{0.49\linewidth}
\center{\includegraphics[width=1.0\linewidth]{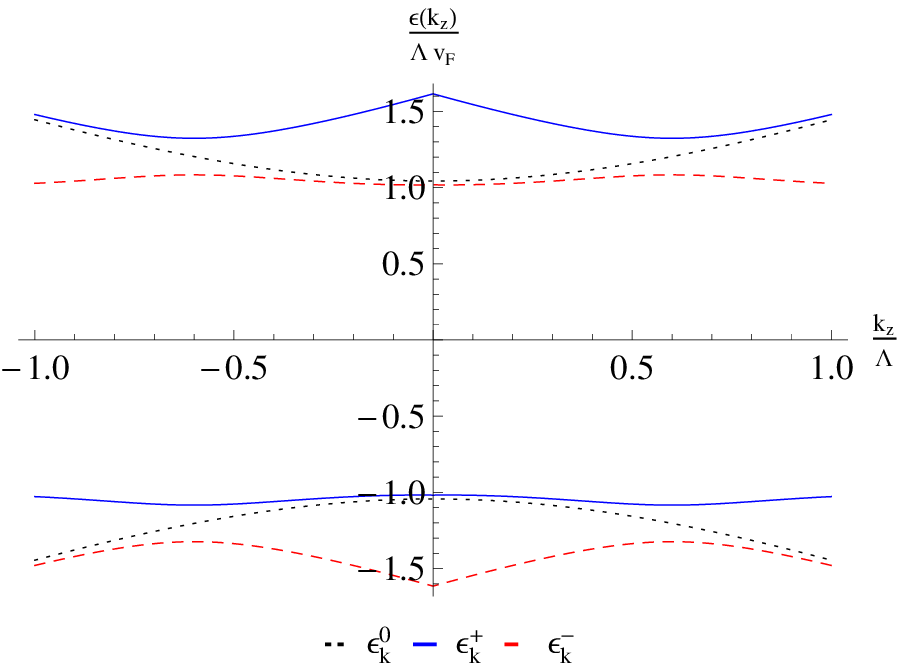} \\ (a)}
\end{minipage}
\hfill
\begin{minipage}[h!]{0.49\linewidth}
\center{\includegraphics[width=1.0\linewidth]{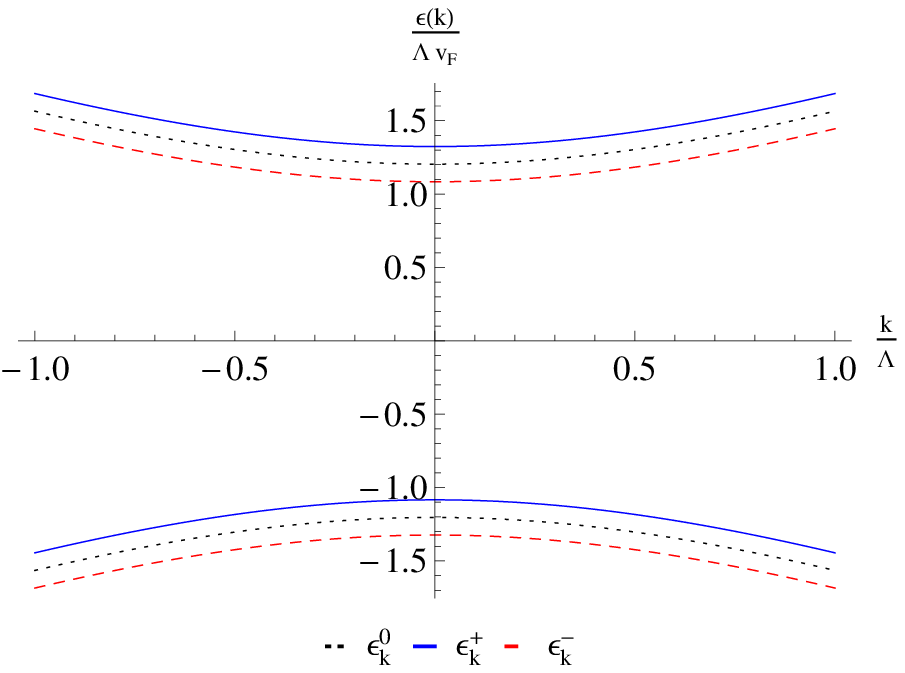} \\ (b)}
\end{minipage}
\caption{(Color online) Quasiparticle energy spectrum in the first order of perturbation theory for the chiral shift $b_{0}=0.6\times \Lambda$ as a function of: (a) $k_z$ with $k=0$, (b) $k$ with $k_z=b_0$. Solid blue lines correspond to $\epsilon_{k}^{+}$, dashed red lines to $\epsilon_{k}^{-}$, and dotted black lines to the zero order of perturbation theory.}
\label{fig:Spectrum_All}
\end{figure*}

In the previous subsection, we have found $\Delta_1$ and $\Delta$. Let us determine the energy spectrum of the system with the dynamically
generated $\Delta_1$ and $\Delta$. Assuming without any loss of generality that $\mathbf{b}_0$ points in the $+z$ direction and performing chiral transformation Eq.~(\ref{chiral-transformation}), we can
rewrite the Hamiltonian of the system as follows:
\begin{eqnarray}
H = v_F\gamma_0(\gamma_x k_x+\gamma_y k_y)-iv_F\gamma_0\gamma_z \partial_z + \gamma_0\Delta +
\gamma_0\Delta_1 e^{2ib_0z\gamma_5}.
\label{Spectrum-Hamiltonian}
\end{eqnarray}
The last term is periodic and has a small amplitude, so we have a standard situation similar to the model of nearly free electron in the solid-state physics. Thus, the quasiparticle energy zones splits into additional zones near the boundaries of a new Brillouin zone. According to Ref.~\cite{Ziman}, we can write the quasiparticle energy spectrum in the first order of perturbation
theory ($\Delta_1\ll\Delta$), i.e.
\begin{eqnarray}
\hspace{-4mm}&&\epsilon^{0}_{k} = \pm \sqrt{v_F^2\mathbf{k}_{\perp}^2+v_F^2k_z^2+\Delta^2}, \nonumber \\
\hspace{-4mm}&&\epsilon_{k}^{\pm} = \frac{\epsilon^{0}_{k}+\epsilon^{0}_{k-K}}{2} \pm \frac{1}{2} \sqrt{\left(\epsilon^{0}_{k}-\epsilon^{0}_{k-K}\right)^2+4U_{K}U_{-K}},
\label{Spectrum-E}
\end{eqnarray}
where $K=2b_0$ is the inverse lattice vector of a new Brillouin zone and,
\begin{eqnarray}
\hspace{-4mm}&&U_{K}=\frac{b_0}{\pi}\int^{\frac{\pi}{2b_0}}_{-\frac{\pi}{2b_0}} dz e^{-iKz} \gamma_0\Delta_1 e^{2ib_0z\gamma_5} = \frac{-4\Delta_1 b_0 \gamma_0 }
{\pi} \sin{\left( \frac{K\pi}{2b_0} \right)} \left( P_{+}\frac{1}{K-2b_0} +P_{-}\frac{1}{K+2b_0} \right) = \nonumber\\
\hspace{-4mm}&&= -2\Delta_1 \gamma_0  \left( P_{+} \delta_{K, 2b_0} + P_{-}\delta_{K, -2b_0} \right).
\label{Spectrum-U}
\end{eqnarray}
To plot the quasiparticles energy spectrum we can use the following numerical values
$v_F \approx 3.5\times 10^5\, m/s$, $g=1.2\times g_{cr}$, $\Delta=0.64\, eV$. For the given $g$, the gap in Fig.~\ref{fig:Delta_Surface} is well fitted by
$\Delta_1 = \frac{1 + c_1 y^2 + c_2 y^4}{c_3 + c_4 y^2 + c_5 y^4} \times (\Lambda v_F)$ with fitting parameters
$c_1=7.8, c_2=19.1, c_3=7.4, c_4=106.8, c_5=453.2$, and $y=\frac{b_{0}}{\Lambda}$. We plot energy spectrum in
Fig.~\ref{fig:Spectrum_All}.

\vspace{5mm}

\section{Discussion and Summary}
\label{5}

We have studied the gap generation in Weyl semimetals, by using a model with local Coulomb interaction. We have showed that there is a critical value of
coupling constant $g_{cr}$, which separates the symmetric phase and the phase with broken symmetry. The phase
diagram of the system is displayed in Fig.~\ref{fig:Phase-diagram-LOFF} in the plane of coupling constant and chiral shift parameter. Further, the gap generation in Weyl semimetals-like materials with small bare gap was studied. The non-zero bare gap considerably complicates the analysis,
because the chiral phase in the ansatz for the inverse full fermion propagator cannot be removed by the chiral transformation. The solution in this case is displayed in Fig.~\ref{fig:Delta_Surface}. Obviously, the chiral shift parameter inhibits the gap generation, because the larger $|\mathbf{b_0}|$ the smaller the gap $\Delta_1$. Further,
the quasiparticle energy spectrum was determined, and it is found that the simultaneous presence of gaps $\Delta_1$ and $\Delta e^{-2ib_0z\gamma_5}$ leads to the additional splitting of the quasiparticle energy bands shown in Fig.~\ref{fig:Spectrum_All}.

In the present study, we have analyzed a simple model with two Weyl nodes and a contact four-fermion interaction. In real materials, such
as tellurium, bismuth, and antimony heterostructures, the more realistic Coulomb interaction and the anisotropy should be taken into account. The
corresponding analysis will be done and reported elsewhere. However, we believe that our qualitative results will survive in the case of more
realistic models.

\vskip5mm
\begin{center}
{\bf Acknowledgments}
\end{center}
The author is grateful to E.V.~Gorbar for helpful discussions.

\appendix

\section{Free energy density}
\label{A3}

In this Appendix, we will focus on the derivation of the free energy density. Since the Schwinger-Dyson equation (\ref{CJT-LR-2})
implies that
\begin{eqnarray}
\frac{g}{2}\int d^4r \left(\mathrm{tr}\big[G(r, r)G(r, r)\big] - \mathrm{tr}\big[G(r,r)\big]\mathrm{tr}\big[G(r,r)\big] \right)= -\frac{i}{2}\left(1-S^{-1}G\right).
\label{Hint-CJT}
\end{eqnarray}
The energy density can be rewritten as follows:
\begin{eqnarray}
\hspace{-4mm}&&{\cal E}=i\mathrm{Tr}\left(\mathrm{Ln}G^{-1} + \frac{1}{2}(S^{-1}G-1)\right)= i\int_{-\infty}^{\infty}\frac{d\omega}{2\pi}\mathrm{Tr}\left[\mathrm{Ln}G^{-1}+\frac{1}{2}
(S^{-1}G-1)\right]= \nonumber\\
\hspace{-4mm}&&=i\int_{-\infty}^{\infty}\frac{d\omega}{2\pi}\mathrm{Tr}\left[-\omega\frac{\partial G^{-1}(\omega)}{\partial\omega}G(\omega)+ \frac{1}{2}\left(S^{-1}
(\omega)G(\omega) -1\right)\right].
\label{effective-potential-02}
\end{eqnarray}
Using the relation
\begin{equation}
\frac{\partial iG^{-1}(\omega;\mathbf{r}, \mathbf{r}^\prime)}{\partial\omega}=\delta(\mathbf{r}-\mathbf{r}^\prime),
\label{delta-G}
\end{equation}
we find
\begin{eqnarray}
{\cal E}=i\int_{-\infty}^{\infty}\frac{d\omega}{2\pi} \mathrm{tr}\left[i\omega G(\omega;\mathbf{0})+ \frac{1}{2}\int d^3\mathbf{r}S_0^{-1}(\omega;
\mathbf{r})G(\omega;-\mathbf{r}) \right]-{\cal E}_0,
\label{effective-potential-03}
\end{eqnarray}
where ${\cal E}$ is the energy density. It is convenient to perform the Fourier transformation
\begin{eqnarray}
{\cal E}=i\int\frac{d\omega d^3\mathbf{k}}{(2\pi)^4} \mathrm{tr}\left[i\omega G(\omega;\mathbf{k})+ \frac{1}{2}S^{-1}(\omega;\mathbf{k})G(\omega;
\mathbf{k}) \right]-{\cal E}_0.
\label{effective-potential-phase-01}
\end{eqnarray}
Using Eq.~(\ref{green-function-LR}) and assuming that $\bar{b}=0$ and $\mathbf{b}_0$ points in the $+z$ direction, we obtain
\begin{eqnarray}
\hspace{-4mm}&&{\cal E} = 2i\int\frac{d\omega d^3\mathbf{k}}{(2\pi)^4} \frac{v_F^2k^2+\omega^2}{v_F^2k^2-\omega^2+\Delta^2} - {\cal E}_0 = \nonumber \\
\hspace{-4mm}&&=-\frac{\Lambda^4 v_F}{8\pi^2} \left[\sqrt{\left(\frac{\Delta}{v_F\Lambda}\right)^2+1} \left(2-\left(\frac{\Delta}{v_F\Lambda}\right)^2\right)+ \left(\frac{\Delta}{v_F\Lambda}\right)^4\arcsinh{\left(\frac{v_F\Lambda}{\Delta}\right)} \right] -\tilde{{\cal E}}_0,
\label{effective-potential-phase-02-A}
\end{eqnarray}
where the constant $\tilde{{\cal E}}_0$ can be omitted.

In the case $\bar{b}\neq0$ there are some subtleties in the determination of the energy density of the system. Performing the chiral transformation Eq.~(\ref{chiral-transformation}), one must be careful with the integral boundaries. To account for this fact we can represent Eq.~(\ref{effective-potential-phase-01}) in the terms of left and right-handed parts:
\begin{eqnarray}
{\cal E}=i\int\frac{d\omega d^3\mathbf{k}}{(2\pi)^4} \mathrm{tr}\left[(P_L+P_R)i\omega G(\omega;\mathbf{k})+ (P_L+P_R)\frac{1}{2}S^{-1}(\omega;\mathbf{k})G(\omega;
\mathbf{k}) \right] = {\cal E}_{L}+{\cal E}_{R}.
\label{effective-potential-phase-01-LR-01}
\end{eqnarray}
If we perform the chiral transformation Eq.~(\ref{chiral-transformation}), then we must redefine the limits of integration as follows
\begin{eqnarray}
\hspace{-4mm}&&{\cal E}_{L}: \quad \int_{-\Lambda}^{\Lambda} dk_z\rightarrow \int_{-\Lambda-b_0}^{\Lambda-b_0} dk_z, \nonumber \\
\hspace{-4mm}&&{\cal E}_{R}: \quad \int_{-\Lambda}^{\Lambda} dk_z\rightarrow \int_{-\Lambda+b_0}^{\Lambda+b_0} dk_z.
\label{boundaries-LR-01}
\end{eqnarray}
For the integral over $\omega$, one can use Eq.~(\ref{effective-potential-phase-01-LR-01}) and Eq.~(\ref{green-function-LR}). We have the following typical integral
\begin{eqnarray}
\int\frac{d\omega_{E} d^3\mathbf{k}}{(2\pi)^4} \frac{\omega^4_{E}+\omega_{E}^2A_2+A_1}{(\omega_E^2+W_1)(\omega_E^2+W_2)}=\int\frac{d\omega_{E} d^3\mathbf{k}}{(2\pi)^4} \left(1- \frac{\omega_{E}^2\left(A_2-W_1-W_2\right)+A_1-W_1W_2}{(\omega_E^2+W_1)(\omega_E^2+W_2)}\right),
\label{effective-potential-b-01}
\end{eqnarray}
where $A_1$ and $A_2$ some functions of the dynamical parameters and $\mathbf{k}$ (since they are rather long, we don't present them here). As we are interested in the difference of energy densities, we can neglect the first term in the brackets in the equation above. Further, using Eq.~(\ref{gap-integrals-01}), we can integrate over $\omega_{E}$
\begin{eqnarray}
\hspace{-4mm}&&{\cal E}_{L}=\int_0^{\Lambda} \int_{-\Lambda-b_0}^{\Lambda-b_0} \frac{kdk \,dk_z}{(2\pi)^2} \frac{1}{\sqrt{W_1}+\sqrt{W_2}} \left[A_2(k_z)- W_1-W_2+\frac{A_1(k_z)-W_1W_2}{\sqrt{W_1W_2}}\right], \nonumber \\
\hspace{-4mm}&&{\cal E}_{R}=\int_0^{\Lambda} \int_{-\Lambda+b_0}^{\Lambda+b_0} \frac{kdk \,dk_z}{(2\pi)^2} \frac{1}{\sqrt{W_1}+\sqrt{W_2}} \left[A_2(-k_z)- W_1-W_2+\frac{A_1(-k_z)-W_1W_2}{\sqrt{W_1W_2}}\right],
\label{effective-potential-b-02}
\end{eqnarray}
where $W_{1, 2}=\left(\Delta^2+v_F^2(\mathbf{k}^2+b^2)\right)\mp 2v_Fb\sqrt{\Delta^2+k_z^2v_F^2}$.

\end{document}